\documentclass[a4paper]{jpconf}
\usepackage{graphicx}
\begin{document}
%
%
\title{Unitary and Renormalizable Theory of Higher Derivative Gravity}
%
\author{Gaurav Narain$^1$, Ramesh Anishetty$^2$}
\address{The Institute of Mathematical Sciences, Taramani, Chennai 600113, India}
\ead{gaunarain@imsc.res.in$^1$, ramesha@imsc.res.in$^2$}
%
%
\begin{abstract}
In 3+1 space-time dimensions, fourth order derivative gravity is 
perturbatively renormalizable. Here it is shown that it describes a unitary theory
of gravitons (with/without an additional scalar) in a limited coupling 
parameter space which includes standard cosmology. 
The running of gravitational constant which includes 
contribution of graviton is computed. It is shown that generically
Newton's constant vanishes at short distance in this perturbatively 
renormalizable and unitary theory.
\end{abstract}
%

\section{Introduction and Theory}
\label{intro}

Perturbative Quantum Field Theory (QFT) has been very successful in describing 
three of the four known forces of nature. It has been used to construct the
Standard Model of particle physics, which has been tested to a great 
accuracy in accelerator experiments. However when the same 
methods of QFT are applied to the Einstein-Hilbert (EH) theory of
gravity, well known problem emerge, namely the theory is 
plagued with UV divergences. At each order of perturbation theory 
one has to add new counter terms to cancel the existing divergences,
as a result the theory loses predictability. 
This is due to the fact that in 3+1 dimensions 
the Newton's constant has the mass dimension $({\rm Mass})^{-2}$.
Classically EH theory is very successful in describing Cosmology
\cite{Dodelson, Weinbergbook}. It was shown that when one studies 
quantum matter fields on curved background \cite{Utiyama1962}, 
four kind of divergences appear: $\sqrt{g}$, $\sqrt{g}R$, 
$\sqrt{g}R^2$, $\sqrt{g}R_{\mu\nu}R^{\mu\nu}$. This was a first hint
that perhaps if when one augments the EH theory with $4$-th order 
metric derivative terms, one might witness UV renormalizabilty. 
Indeed it was shown in \cite{Stelle} that the $4$-th order 
metric derivative gravity is UV renormalizable using $4-\epsilon$ 
dimensional regularization scheme \cite{Diagram}. The most general action that 
includes all terms up to $4$-th order metric derivative terms
is given by,
\begin{equation}
\label{eq:hdgact}
A
= \int \frac{{\rm d}^4x \sqrt{-g}}{16 \pi G} \left[
2 \Lambda -R + \frac{\omega R^2}{6 M^2}
- \frac{R_{\mu\nu}R^{\mu\nu} - 
\frac{1}{3}R^2}{M^2}
\right] \, ,
\end{equation}
where $R$ is the Ricci scalar and $R_{\mu\nu}$ is the 
Ricci tensor. The last term is proportional to the square 
of Weyl tensor in 3+1 dimensions up to an Euler characteristic.
Here $\omega$ is dimensionless and $M$ has dimensions of mass. 

Doing perturbations around the flat background 
$g_{\mu\nu} = \eta_{\mu\nu} + h_{\mu\nu}$ 
($\eta_{\mu\nu} = \{1,-1,-1,-1 \}$) one obtains the 
propagator of the theory \cite{Salam, Julve}. Here we will 
consider the action $A$ for $\Lambda=0$ in the 
Landau gauge $\partial^{\mu} h_{\rho\mu}=0$, 
where the Feynman propagator of theory in momentum space 
is given by,
\begin{equation}
\label{eq:grav_prop1}
D_{\mu\nu, \alpha\beta} =
 \frac{i \, 16 \pi G}{(2 \pi)^4} \cdot \Biggl[
\frac{(2P_2 - P_s)_{\mu\nu, \alpha\beta}}{q^2 + i \, \epsilon}
+ \frac{(P_s)_{\mu\nu, \alpha\beta}}{q^2 - M^2/\omega + i \epsilon}
- \frac{2 \, (P_2)_{\mu\nu, \alpha\beta}}{q^2 - M^2+ i \epsilon}
\Biggr] \, ,
\end{equation}
where $q$ is the momentum of fluctuating field $h_{\mu\nu}$. Various 
spin projectors are  
$(P_2)_{\mu\nu, \alpha \beta} = 
\frac{1}{2} \left[ T_{\mu\alpha} T_{\nu\beta} + 
T_{\mu\beta}T_{\nu\alpha} \right] 
- \frac{1}{3} T_{\mu\nu}T_{\alpha\beta} $, 
$(P_s)_{\mu\nu, \alpha \beta} = 
\frac{1}{3} T_{\mu\nu} \, T_{\alpha \beta}$, 
where $T_{\mu\nu}=\eta_{\mu\nu} - q_{\mu}q_{\nu}/q^2$.
The first term of Eq. (\ref{eq:grav_prop1}) is the massless spin-2
graviton with two degrees of freedom, the second term 
is the scalar (Riccion) of mass $M/\sqrt{\omega}$ and the last
term is that of massive spin-2 with mass $M$ ($M$-mode).
The last term arises due to the presence of $R_{\mu\nu}R^{\mu\nu}$ term
in the action. If we only had $F(R)$ type of theory, this term will be 
absent. From the propagator Eq. (\ref{eq:grav_prop1}) we note that the residues
at the pole for graviton and Riccion are positive, while for the $M$-mode
it is negative. The $M$-mode renders the theory non-unitary.

We now study the issue of unitarity of higher derivative gravity
action given in Eq. (\ref{eq:hdgact}). The full $S$-matrix of theory
involve gravitons, Riccions and $M$-mode as the external legs.
Lets consider a subpart of this $S$-matrix which involve 
only Riccion and graviton as external legs, which means that we are 
considering scattering process which involve only Riccion and 
graviton but no $M$-mode as external leg. In this subpart of 
$S$-matrix we ask whether this under some conditions 
remains unitary? In any such scattering process, $M$-mode
can appear as an intermediate state, which is not 
present in our subpart of $S$-matrix. For intermediate energies 
$\mu$ less than $M$ it cannot occur anyway. In renormalized 
quantum field theory these mass parameter are also 
energy dependent. Hence if $M(\mu)$ is always greater 
than $\mu$ then it cannot occur at any intermediate energy,
consequently the subpart of $S$-matrix can also satisfy unitarity. In this 
scenario the physical amplitudes do get contributions from 
the $M$-mode but not from the imaginary part of $M$-mode
{\it i.e.} Cutkosky cut corresponding $M$-mode vanish
identically. Thus in the renormalized theory if $M^2(\mu)/\mu^2>1$
is satisfied, then we can have unitary theory of only gravitons and 
Riccions \cite{Narain}. In the following we will see how
this can be realized.

\section{Renormalization Group Analysis}
\label{RGanalysis}

We choose to parameterize our gravity action as in Eq. (\ref{eq:hdgact}),
so that in the path-integral $G$ effectively plays the same
role as $\hbar$ (in the absence of matter fields) 
{\it i.e.} the loop expansion is same as perturbation 
theory in small $G$. To one-loop beta function of couplings have been computed 
using the Schwinger-Dewitt technique \cite{Dewitt}. In 
\cite{Fradkin,avramidi} this was done using dimensional 
regularization in $4-\epsilon$ space-time dimensions. 
In the Landau gauge the beta function are the following \cite{avramidibook}:
\begin{eqnarray}
\label{eq:betaM2G}
&& \frac{{\rm d}}{{\rm d} t} \left( \frac{1}{M^2 G} \right)
= - \frac{133}{10 \pi} \, ,
\\
\label{eq:beta_wM2G}
&& \frac{{\rm d}}{{\rm d} t} \left(
\frac{\omega}{M^2 G} \right) =
\frac{5}{3 \pi} \left( \omega^2
+ 3 \omega + \frac{1}{2} \right) \, ,
\\
\label{eq:betaG}
&& \frac{{\rm d}}{{\rm d} t} \left(\frac{1}{G} \right)
= \frac{5M^2}{3 \pi} \left( \omega
- \frac{7}{40 \omega} \right) \, ,
\\
\label{eq:betalamG}
&& \frac{{\rm d}}{{\rm d} t} \left(
\frac{2 \Lambda}{G} \right) = 
\frac{M^4}{2 \pi} \left(5 + \frac{1}{\omega^2}
\right)
- \frac{4M^2 \Lambda}{3 \pi} \left(
14 + \frac{1}{\omega}
\right) \, ,
\end{eqnarray}
where $t= \ln (\mu/\mu_0)$ and all $rhs$ contains
the leading contribution in $G$ ($M^2 G$ is also taken
to be small), with higher powers coming from 
higher loops being neglected. 
Using Eqs. (\ref{eq:betaM2G} and \ref{eq:beta_wM2G}) we solve
for the running of $\omega$,
\begin{equation}
\label{eq:beta_w}
\frac{{\rm d} \omega}{{\rm d} t} 
= \frac{5 M^2 G}{3 \pi} 
\left(
\omega^2 + \frac{549}{50} \omega + \frac{1}{2} 
\right) 
= \frac{5 M^2 G}{3 \pi}  (\omega + \omega_1)(\omega + \omega_2) \, ,
\end{equation}
where $\omega_1= 0.0457$ and $\omega_2 = 10.9343$. 
On studying Eq. (\ref{eq:beta_w}) we find that the RG flow of 
$\omega$ has two fixed point: $-\omega_1$ and $-\omega_2$,
with former being repulsive and later attractive under the 
UV evolution or increasing $t$. We realize that both these 
fixed points lie in the unphysical domain. The last equality 
of Eq. (\ref{eq:beta_w}) tells that $rhs$ is always positive for $\omega>0$.
This means that $\omega$ is a monotonic increasing function of $t$
and vice-versa. Eq. (\ref{eq:betaM2G}) is easily integrated to express
the flow of $M^2G$ in terms of $t$. This is then plugged in 
Eq. (\ref{eq:beta_w}) to obtain,
\begin{equation}
\label{eq:tw}
t= T
\Biggl[
1- 
\left( \frac{\omega + \omega_2}{\omega + \omega_1}
\cdot 
\frac{\omega_0 + \omega_1}{\omega_0 + \omega_2} \right)^{\alpha}
\Biggr] \, ,
\end{equation}
where $T=10 \pi/(133 M_0^2 G_0)$ 
and $\alpha= 399/50(\omega_2 -\omega_1)$, with
subscript $0$ meaning that the coupling parameters are evaluated at $t=0$ or 
$\mu = \mu_0$. As $\omega$ takes value between zero 
and infinity, this translates using Eq. (\ref{eq:tw}) in to a
range for $t$. We note that $t$ takes a minimum value 
for $\omega=0$ while a maximum value for $\omega=\infty$.
\begin{equation}
\label{eq:tmax_min}
\frac{t_{min}}{T} \equiv 
1 - \left(\frac{\omega_2}{\omega_1} 
\frac{\omega_0 + \omega_1}
{\omega_0 + \omega_2}
\right)^{\alpha}
\leq \frac{t}{T}
< 
1 - \left(
\frac{\omega_0 + \omega_1}
{\omega_0 + \omega_2}
\right)^{\alpha}
\equiv \frac{t_{max}}{T} \, .
\end{equation}

Using the evolution equation for $\omega$, one can transform 
any evolution of coupling in $t$ space to $\omega$ space.
This allows analytical expressions for the flow of other couplings.
Using Eqs. (\ref{eq:betaG} and \ref{eq:beta_w}) we get,
\begin{equation}
\label{eq:betalogG}
\frac{{\rm d} \ln G}{{\rm d} \omega} 
= - \frac{\omega - \frac{7}{40 \omega} }
{(\omega + \omega_1) (\omega + \omega_2)} \, ,
\hspace{5mm}
\frac{G}{G_0} = \frac{\omega_0}{\omega} \cdot 
\left(
\frac{1 + \omega_1/\omega }
{1 + \omega_1/\omega_0 }\right)^{A1}
\left(
\frac{1 + \omega_2/\omega }
{1 + \omega_2/\omega_0 }\right)^{A2} \, ,
\end{equation}
where the first equation expresses the running of $\ln G$ in 
$\omega$ space, while in second equation we write its solution.
The first equation tells that during the RG evolution $G$ gets extremised 
at $\omega = \sqrt{7/40}$, taking the second derivative
it is shown that this point is a maxima. We choose this 
point to be our reference point $\mu_0$ or $t=0$ and integrate
to obtain the second equation,
where $A1=-0.3473$ and $A2= -1.0027$. 
Eq. (\ref{eq:betalogG}) show that 
for large $\omega$ or $t$, $G \sim 1/\omega$, thereby
vanishing for large $t$ \cite{Donoghue}, while
for small $\omega$, $G \sim \omega^{7/20}$
reaching a peak at $\omega_0=\sqrt{7/40}$.
Similarly, using Eqs. (\ref{eq:beta_wM2G} and \ref{eq:betaG})
we obtain the running of $M^2/\omega$, which along with
with Eq. (\ref{eq:beta_w}) is integrated to give
$M^2/\omega = (M_0^2/\omega_0)
\left(
(1 + \omega_1/\omega )/(1 + \omega_1/\omega_0 )\right)^{B1} 
\left(
(1 + \omega_2/\omega )(1 + \omega_2/\omega_0 )\right)^{B2}$,
where $B1=1.0802$ and $B2= 0.2698$.
This tells that as $\omega \to \infty$, the mass of the 
$M$-mode also goes to infinity, which means that ultimately 
it is decoupled from the theory. But does it goes to infinity 
quick enough so that it is never encountered in the theory,
is the question we ask next? To answer this we now analyze $M^2/\mu^2$.
It is instructive to note that from
Eqs. (\ref{eq:betaM2G}, \ref{eq:betaG} and \ref{eq:beta_w}) we obtain
the running of $\ln (M^2/\mu^2)$,
\begin{equation}
\label{eq:beta_M2mu}
\frac{{\rm d}}{{\rm d} \omega} \ln \left( \frac{M^2}{\mu^2} \right)
= \frac{ \left(
\omega + \frac{399}{50} - \frac{7}{40 \omega}
- \frac{6 \pi}{5 M^2 G}
\right) }
{ (\omega + \omega_1)
(\omega + \omega_2)}
\, .
\end{equation}
This shows that $M^2/\mu^2$ reaches a minima for 
$\omega = \omega_*$ given by 
$
\left(\omega_* + 399/50 - 7/40 \omega_* \right)
= 6 \pi/(5 M_*^2 G_*)$.
Hence by demanding
$M^2_*/\mu^2_* = (6 \pi/5 \mu^2_* G_*)/
(\omega_* + 399/50 - 7/40 \omega_*)> 1$,
we make the $M$-mode not realizable
in the physical GR sector of the theory.
This inequality is easily achievable by choosing $\mu^2_* G_*$ 
appropriately. Perturbative loop expansion requires that 
$M^2G$ is small. Therefore $M$ is a sub-Planckian mass,
yet the running mass as dictated by interactions makes it
physically not realizable even in post Planckian regime.

One can do a similar analysis to answer the question whether
Riccion is realizable or not? For this we study the RG evolution
of $M^2/(\omega \mu^2)$.
From Eqs. (\ref{eq:beta_wM2G}, \ref{eq:betaG}
and \ref{eq:beta_w}) we obtain
$ {\rm d} \ln \left(M^2/\omega \mu^2 \right) /{\rm d}\omega
= - \left(3 + 27/40 \omega + 6 \pi/5 M^2 G \right)/
(\omega + \omega_1)(\omega + \omega_2)$,
showing that the Riccion mass relative to $\mu$
decreases monotonically. By a suitable choice we can make the
Riccion to be physically realizable or not. So we conclude that 
there exists unitary physical subspace only 
with the gravitons or along with Riccions.

To make the running of $\Lambda$ to zero we add
two spin-$\frac{1}{2}$ Dirac fields (for detail see \cite{Narain})
with mass $(5/4)^{1/4}M$ and $M/\sqrt{2\omega}$ \cite{Gorbar}
(these additional fermionic fields can be interpreted as ghosts
normalizing the functional integral). This assures that 
if initially $\Lambda=0$, then throughout 
the RG flow $\Lambda$ will remain zero. The affect of adding the 
two Dirac fields is just to shift $\omega_0$, $\omega_1$, 
$\omega_2$ and $\omega_*$, but all conclusions remains unaltered.

In arbitrary gauge the beta function $1/M^2G$ and $\omega/M^2G$ remains
unaltered, which means that running of $\omega$ and
two fixed points $\omega_1$ and $\omega_2$
remains same. However the beta function of $1/G$ gets modified. In a general
gauge it is
${\rm d} \ln G/{\rm d} t
= - \left(5M^2 G/3\pi \right) \left(
\omega + a + b/\omega \right)$, 
where $a$ and $b$ depend on the gauge-fixing parameters 
used, while the leading term is gauge invariant. This means 
that $\omega_0$ shifts. Doing the 
same analysis as before, we write the running of $G$ in terms 
of $\omega$, from which we note that for large $\omega$ 
or $t$, $G \sim 1/\omega$ is a gauge invariant result, while
for small $\omega$, the rate at which $G$ vanishes depend 
on $a$ and $b$.


\section{Discussion}
\label{discuss}

By doing the one-loop analysis we have found that there is no 
solution for $\omega$ from Eq. (\ref{eq:tw}) if $t>t_{max}$. This 
is because $\omega$ reaches its maximum value infinity. In the 
other extreme when $\omega=0$, $t$ reaches its minimum value $t_{min}$.
For $t<t_{min}$, $\omega$ become negative {\it i.e.}
$M^2/\omega$ the Riccion mass square becomes negative
signaling the instability of the vacuum. 


Finally we conclude that the action $A$, Eq. (\ref{eq:hdgact}) 
describes a perturbative quantum 
gravity as self consistent, renormalizable and unitary theory 
of gravitons and the curvature cannot become
singular, in particular it cannot fluctuate wildly at sub-Planckian 
\cite{Starobinsky} or post Planckian regimes
consistent with known cosmology. 
Its predictions asymptotically beyond Planck scale needs 
to be investigated further.

\section*{References}
%

%
%
\end{document}